\documentclass[twocolumn,showpacs,preprintnumbers,amsmath,amssymb]{revtex4}

\usepackage{graphicx}
\usepackage{dcolumn}
\usepackage{bm}
\usepackage{hyperref}

\pdfoutput=1

\hypersetup{pdfauthor={some author},pdftitle={eye-catching title}}
\begin{document}

\preprint{APS/123-QED}

\title{Novel Isotope Effects on the Pairing Pseudogap in High-$T_{c}$ Cuprates:
Evidences for Polaronic Metal and Precursor BCS-Like Pairing of
Large Polarons}

\author{S. Dzhumanov}
\email{dzhumanov@rambler.ru} \affiliation{%
Institute of Nuclear Physics, 100214, Ulughbek, Tashkent,
Uzbekistan}
\author{O.K. Ganiev}
\affiliation{%
Institute of Nuclear Physics, 100214, Ulughbek, Tashkent,
Uzbekistan}
\author{Sh.S. Djumanov}
\affiliation{%
Institute of Nuclear Physics, 100214, Ulughbek, Tashkent,
Uzbekistan}

\begin{abstract}
We have studied the novel isotope effects on the pairing pseudogap
in underdoped and optimally doped cuprates within the large-polaron
model and two non-standard BCS-like approaches. We have shown that
in the intermediate-coupling regime the precursor pairing of large
polarons occurs at a mean-field temperature $T^{\ast}>T_{c}$ and the
near-absent, sizable and very large oxygen and copper isotope
effects on $T^{\ast}$ exist in cuprates with small and large Fermi
surfaces. Our results for $T^{\ast}$, isotope shifts and exponents
in slightly underdoped and optimally doped cuprates are in
quantitative agreement with existing experiments and explain the
discrepancy between various experiments.
\end{abstract}

\maketitle

Superconductivity in ordinary and polaronic metals is the result of
two distinct quantum phenomena (as argued in Refs.\cite{1,2,3,4,5}):
pairing of charge carriers at a characteristic temperature
$T^{\ast}$ and condensation of bound pairs into a superfluid state
at the superconducting (SC) transition temperature $T_{c}$. In
conventional superconductors, pairing of electrons and superfluidity
of Cooper pairs occur simultaneously at $T^{\ast}=T_{c}$. In
high-$T_c$ cuprates, however, the pairing of charge carriers may
occur at a higher temperature $T^{\ast}$ than the $T_{c}$ at which
the pre-formed Cooper pairs condense into a superfluid (or SC) state
\cite{1,2,3}. As a consequence, below $T^{\ast}$ a pseudogap (PG)
appears in cuprates, as seen in many experiments (NMR, ARPES,
optical conductivity, specific heat, etc.) (see Refs.\cite{6,7,8}
for a review). Most of the theories proposed for the high-$T_{c}$
cuprates are based on different electronic mechanisms of pairing and
ingnore effects arising from the electron-phonon interactions.
However, a number of experiments revealing the peculiar isotope
effects on $T_{c}$, $T^{\ast}$ and other physical quantities in
cuprates \cite{9,10,11,12,13,14,15} have shown that the
electron-phonon interactions play a major role in these materials.
In particular, some experiments showed that the oxygen and copper
isotope effects on $T^{\ast}$  in \rm{Y}- and \rm{La}- based
cuprates are absent or very small \cite{9,14} and sizable
\cite{9,13,14}. While other experiments revealed the presence of
large negative oxygen and copper isotope effects on $T^{\ast}$ in
\rm{Ho}-based cuprates \cite{11,13}. The origin of the novel isotope
effects on $T^{\ast}$ in cuprates has especially been the subject of
controversy \cite{9,10,11}. Although some theories \cite{16,17,18}
were used to describe the oxygen isotope effect on $T^{\ast}$ in
high-$T_c$ cuprates, they cannot give an answer to the puzzling
question why such an isotope effect is sizable or very large in some
cuprates and absent in others. Also, there is no explanation of the
observed copper isotope effect on $T^{\ast}$ in cuprates.

In this Letter, we address the above issues of the isotope effects
on the pairing PG in the cuprates by proposing two BCS-based
approaches extended to the intermediate-coupling regime, which allow
to describe the precursor pairing of large polarons above $T_{c}$ in
underdoped and optimally doped cuprates with small and large Fermi
surfaces. We derive new expressions for the mean-field pairing
temperature $T^{\ast}$ and the isotope exponents
$\alpha_{T^{\ast}}$. Remarkably, our results for $T^{\ast}$, isotope
shifts and exponents in slightly underdoped and optimally doped
cuprates are in quantitative agreement with existing experiments and
resolve the controversy between various experiments. We also make
quantitative predictions for the isotope effects on $T^{\ast}$ in
the deeply underdoped cuprates.

In conventional superconductors the mass of free electrons is
independent of the ionic mass $M$. In contrast, the charge carriers
in polar materials are self-trapped and form large polarons
(quasiparticles dressed by lattice distortions) \cite{19}. In the
polaronic model, the mass of polarons $m_{p}$ depends on the
longitudinal-optical (LO) phonon-frequency $\omega_{LO}$ which in
turn depends on $M$. The observed effective mass of charge carriers
in cuprates is about $2m_{e}$ (where $m_{e}$ is the free electron
mass) \cite{20} and the relatively small binding energies
$E_{p}=0.06 eV$ and $E_{p}=0.12 eV$ of the polaron are observed in
optimally doped and underdoped cuprates, respectively \cite{21}.
These and other experiments \cite{20,22} prove that the charge
carriers in cuprates are large polarons.

Here we consider the real physical situations in cuprates, namely,
the cases of small and large Fermi surfaces. In slightly underdoped
and optimally doped cuprates with large Fermi surfaces, the new
situation arises when the polaronic effect exists and the attractive
interaction mechanism (e.g., due to exchange of static and dynamic
phonons) between the carriers operating in the energy range
$\{-(E_{p}+\hbar\omega_{LO}), (E_{p}+\hbar\omega_{LO})\}$ is more
effective than in the simple BCS picture. The energies $\varepsilon$
of polarons are measured from the polaronic Fermi energy
$\varepsilon_{F}$. It was argued \cite{1,2,3} (see also Ref.
\cite{23}) that the extension of the BCS theory to the intermediate
coupling regime describes the precursor pairing of carriers above
$T_{c}$. At $\varepsilon_{F}>E_{p}+\hbar\omega_{LO}>>k_{B}T^{\ast}$,
using a similar BCS-based approach, we obtain a new and more general
expression for the mean-field pairing temperature $T^{\ast}$:
\begin{eqnarray}\label{Eq.1}
k_{B}T^{\ast}=1.14(E_{p}+\hbar\omega_{LO})\exp[-1/\lambda^{\ast}],
\end{eqnarray}
where $\lambda^{\ast}=N_{p}(\varepsilon_{F})\tilde{V}_{p}$ is the
BCS-like coupling constant, $N_{p}(\varepsilon_{F})$ is the
polaronic density of states (DOS), $\tilde{V}_{p}=V_{ph}-V_{c}$ is
the effective polaron-polaron interaction potential, which has both
the attractive electron-phonon interaction part $V_{ph}$ and the
screened repulsive Coulomb interaction part
$\tilde{V_{c}}=V_{c}/\left[1+N_{p}(\varepsilon_{F})V_{c}\ln{(\zeta_{c}/(E_{p}+\hbar\omega_{LO}))}\right]$,
$\zeta_{c}$ is the cut-off for the Coulomb interaction $V_c$,
$\zeta_{c}>E_{p}+\hbar\omega_{LO}$. The usual BCS approximation as
the particular case is recovered in the weak coupling regime (i.e.
in the absence of polaronic effects, $E_{p}=0$) and the prefactor in
Eq.(\ref{Eq.1}) is replaced by Debye energy $\hbar\omega_{D}$. Here
we show that the polaronic effects change significantly the simple
BCS picture and are responsible for the novel isotope effects on
$T^{\ast}$. In the large polaron theory, $m_{p}$, $E_{p}$ and
$\varepsilon_{F}=\hbar^{2}(3\pi^{2}n)^{2/3}/2m_{p}$ (where $n$ is
the concentration of large polarons) depend on the Fr\"{o}hlich-type
electron-phonon coupling constant $\alpha_F$ which in turn depends
on the masses $M(=M_{O}$ or $M_{Cu})$ and $M^{'}(=M_{Cu}$ or
$M_{O})$ of the oxygen \rm{O} and copper \rm{Cu} atoms in cuprates:
\begin{eqnarray}
\alpha_{F}=\frac{e^2}{2\hbar\omega_{LO}}\left[\frac{1}{\varepsilon_{\infty}}-\frac{1}{\varepsilon_{0}}\right]\left(\frac{2m^{\ast}\omega_{LO}}{\hbar}\right)^{1/2},
\end{eqnarray}
where
$\omega_{LO}\simeq\left(2\beta\left(\frac{1}{M}+\frac{1}{M^{'}}\right)\right)^{1/2}$,
$\beta$ is a force constant of the lattice, $\varepsilon_{\infty}$
and $\varepsilon_{0}$ are the high frequency and static dielectric
constants, respectively, $m^{\ast}$ is the effective mass of a
carrier in the absence of the electron-phonon interaction. In the
intermediate electron-phonon coupling regime the mass and binding
energy of a large polaron are given by
$m_{p}=m^{\ast}(1+\alpha_{F}/6)$ and
$E_{p}=\alpha_{F}\hbar\omega_{LO}$ \cite{19}

The polaronic DOS can be approximated in a simple form
\begin{eqnarray}
N_{p}(\varepsilon_{F})= \left\{ \begin{array}{ll}
1/\varepsilon_{F} & \textrm{for\quad $0<\varepsilon\leq\varepsilon_{F}$}\\
0 & \qquad \textrm{otherwise}.
\end{array} \right.
\end{eqnarray}
From Eq.(\ref{Eq.1}), the exponent of the isotope effect on
$T^{\ast}$ is defined as
\begin{eqnarray}\label{Eq.4}
\alpha_{T^{\ast}}=-\frac{d\ln{T^{\ast}}}{d\ln{M}}.
\end{eqnarray}
Using the above expressions for $m_{p}$, $E_{p}$ and
$N_{p}(\varepsilon_{F})$, the Eqs. (\ref{Eq.1}) and (\ref{Eq.4}) can
be written as
\begin{eqnarray}\label{Eq.5}
k_{B}T^{\ast}=1.14A\mu^{-1/4}\left(1+a\mu^{-1/4}\right)\exp\left[-1/\lambda^{\ast}(\mu)\right]
\end{eqnarray}
and
\begin{widetext}
\begin{eqnarray}\label{Eq.6}
\alpha_{T^{\ast}}=\frac{1}{4(1+M/M^{'})}\left\{1+\frac{a\mu^{-1/4}}{1+a\mu^{-1/4}}-
\frac{1}{(\lambda^{\ast}(\mu))^{2}}
\left(\lambda_{ph}b\mu^{1/4}-\frac{\lambda_{c}b\mu^{1/4}}{U_{c}(\mu)}\right.\right.\nonumber\\
\left.\left.+\frac{\lambda_{c}^{2}(1+b\mu^{1/4})}{U^{2}_{c}(\mu)}\left[b\mu^{1/4}\ln{B_{c}(\mu)}+(1+b\mu^{1/4})\left(1+\frac{a\mu^{-1/4}}{1+a\mu^{-1/4}}\right)\right]\right)\right\},
\end{eqnarray}
\end{widetext}
where
$\lambda^{\ast}(\mu)=\lambda_{ph}(1+b\mu^{1/4})-\lambda_{c}(1+b\mu^{1/4})/U_{c}(\mu)$,
$U_{c}(\mu)=1+\lambda_{c}(1+b\mu^{1/4})\ln{B_{c}(\mu)}$,
$\lambda_{ph}=\left[2m^{\ast}/\hbar^{2}(3\pi^{2}n)^{2/3}\right]V_{ph}$,
$\lambda_{c}=\left[2m^{\ast}/\hbar^{2}(3\pi^{2}n)^{2/3}\right]V_{c}$,
$B_{c}(\mu)=\zeta_{c}/A\mu^{-1/4}\left(1+a\mu^{-1/4}\right)$,
$A=\frac{e^{2}}{\tilde{\varepsilon}}\sqrt{\frac{m^{\ast}}{2\hbar}}(2\beta)^{1/4}$,
$a=\hbar\tilde{\varepsilon}\sqrt{\frac{2\hbar}{m^{\ast}}}(2\beta)^{1/4}/e^2$,
$b=1/6a$,
$\tilde{\varepsilon}=\varepsilon_{\infty}/(1-\varepsilon_{\infty}/\varepsilon_{0})$,
$\mu=MM^{'}/(M+M^{'})$ is the reduced mass of ions.
\par Equations (\ref{Eq.5}) and (\ref{Eq.6}) allow us to calculate the PG
temperature $T^{\ast}$ and the exponents $\alpha^{O}_{T^{\ast}}$ and
$\alpha^{Cu}_{T^{\ast}}$ of the oxygen and copper isotope effects on
$T^{\ast}$.
\begin{figure}
\includegraphics[width=0.51\textwidth]{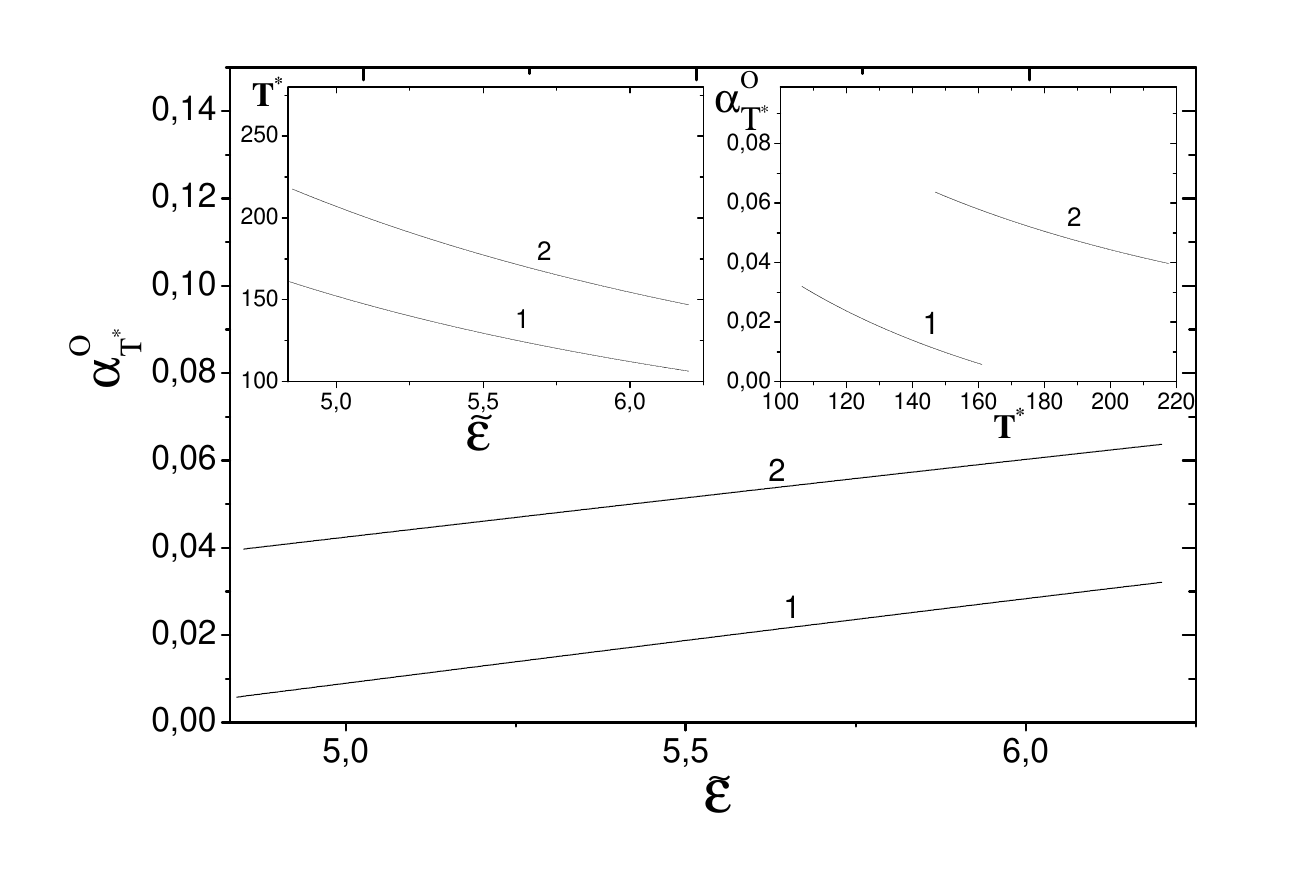}
\caption{\label{fig1:epsart}\footnotesize Variation of
$\alpha^{O}_{T^{\ast}}$ as a function of $\tilde{\varepsilon}$ for
two sets of parameters:(1) $\lambda_{ph}=0.297$,
$\lambda_{c}=0.077$, $n=0.92\cdot10^{21} cm^{-3}$ and (2)
$\lambda_{ph}=0.313$, $\lambda_{c}=0.067$, $n=0.94\cdot10^{21}
cm^{-3}$. The insets show the dependences
$T^{\ast}(\tilde{\varepsilon})$ and
$\alpha^{O}_{T^{\ast}}(\tilde{\varepsilon})$ for the same values of
$\lambda_{ph}$, $\lambda_{c}$, $n$.}
\end{figure}
In our numerical calculations, we take $m^{\ast}\simeq m_{e}$
\cite{20}, $\varepsilon_{\infty}=3 - 5$ \cite{20,24},
$\varepsilon_{0}=22 - 30$ \cite{20,24} and $\hbar\omega_{LO}=0.04 -
0.07 eV$ \cite{7,20}, typical values for the cuprates. Then the
values of $\tilde{\varepsilon}$ and $\alpha_{F}$ are
$\tilde{\varepsilon}\simeq3.33 - 6.47$ and $\alpha_F=2.15 - 5.54$
(which correspond to the intermediate electron-phonon coupling
regime). Notice that in discussing the experimental data for isotope
effects on $T^{\ast}$ in cuprates we have taken the fact that the
physical situations (doping levels, dielectric constants,
$T^{\ast}$) in various experiments are rather different. The
magnitude of $\beta$ is kept at the value estimated for the oxygen
and copper unsubstituted compound using a given value of
$\hbar\omega_{LO}$. Since the quantity $\zeta_{c}$ is of the order
of $\varepsilon_{F}$, the logarithm $\ln{B_{c}(\mu)}$ will be small,
so that the Coulomb pseudopotential $\tilde{V}_{c}$ is of the order
of bare Coulomb potential $V_{c}$. The results of numerical
calculations of $T^{\ast}$ and $\alpha_{T^{\ast}}$ according to
Eqs.(\ref{Eq.5}) and (\ref{Eq.6}) at different values of
$\hbar\omega_{LO}$, $\tilde{\varepsilon}$, $n$, $\lambda_{ph}$ and
$\lambda_{c}$ are shown in Figs.1 - 3.
\begin{figure}
\includegraphics[width=0.9\textwidth]{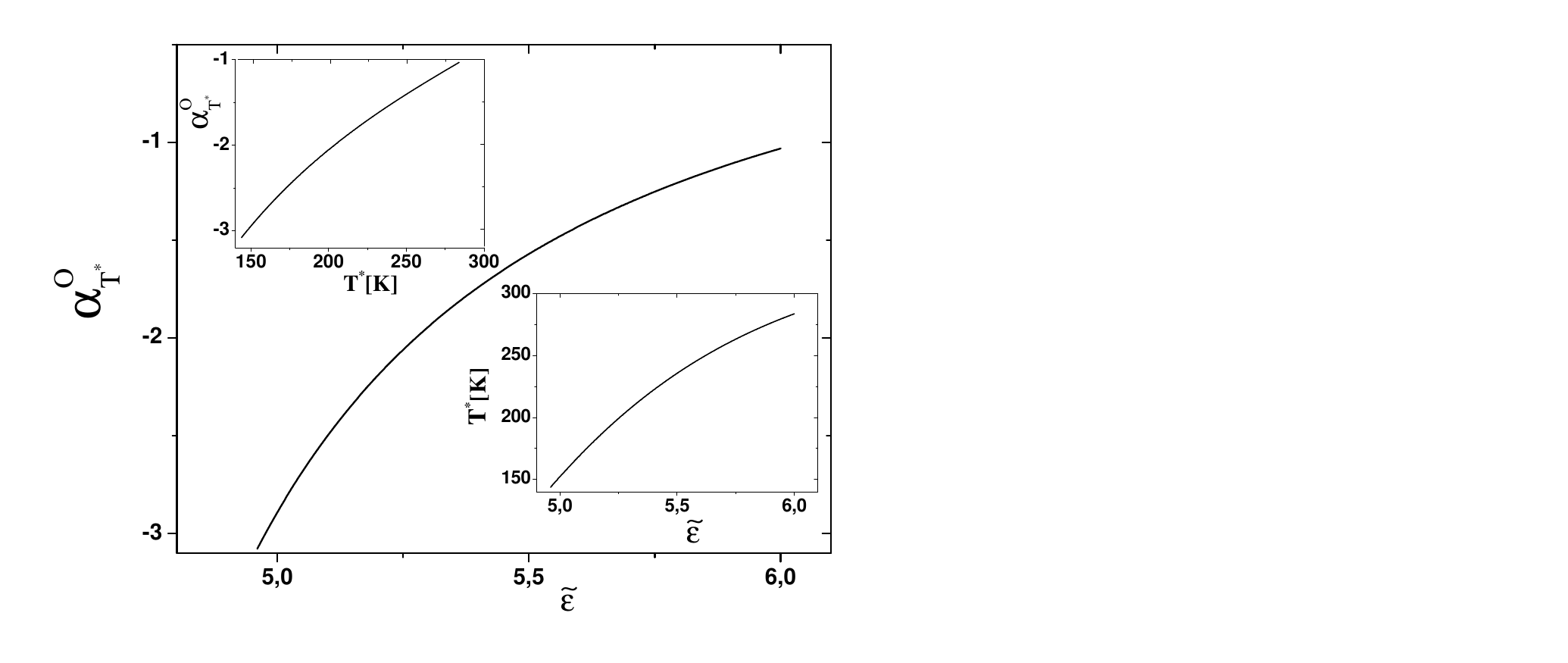}
\caption{\label{fig2:epsart}\footnotesize The dependence of
$\alpha^{O}_{T^{\ast}}$ on $\tilde{\varepsilon}$ for
$\lambda_{ph}=0.975$, $\lambda_{c}=0.820$ and $n=0.88\cdot10^{21}
cm^{-3}$ The insets show the dependences
$T^{\ast}(\tilde{\varepsilon})$ and
$\alpha^{O}_{T^{\ast}}(\tilde{\varepsilon})$ for the same values of
$\lambda_{ph}$, $\lambda_{c}$, $n$.}
\end{figure}
\begin{figure}
\includegraphics[width=0.51\textwidth]{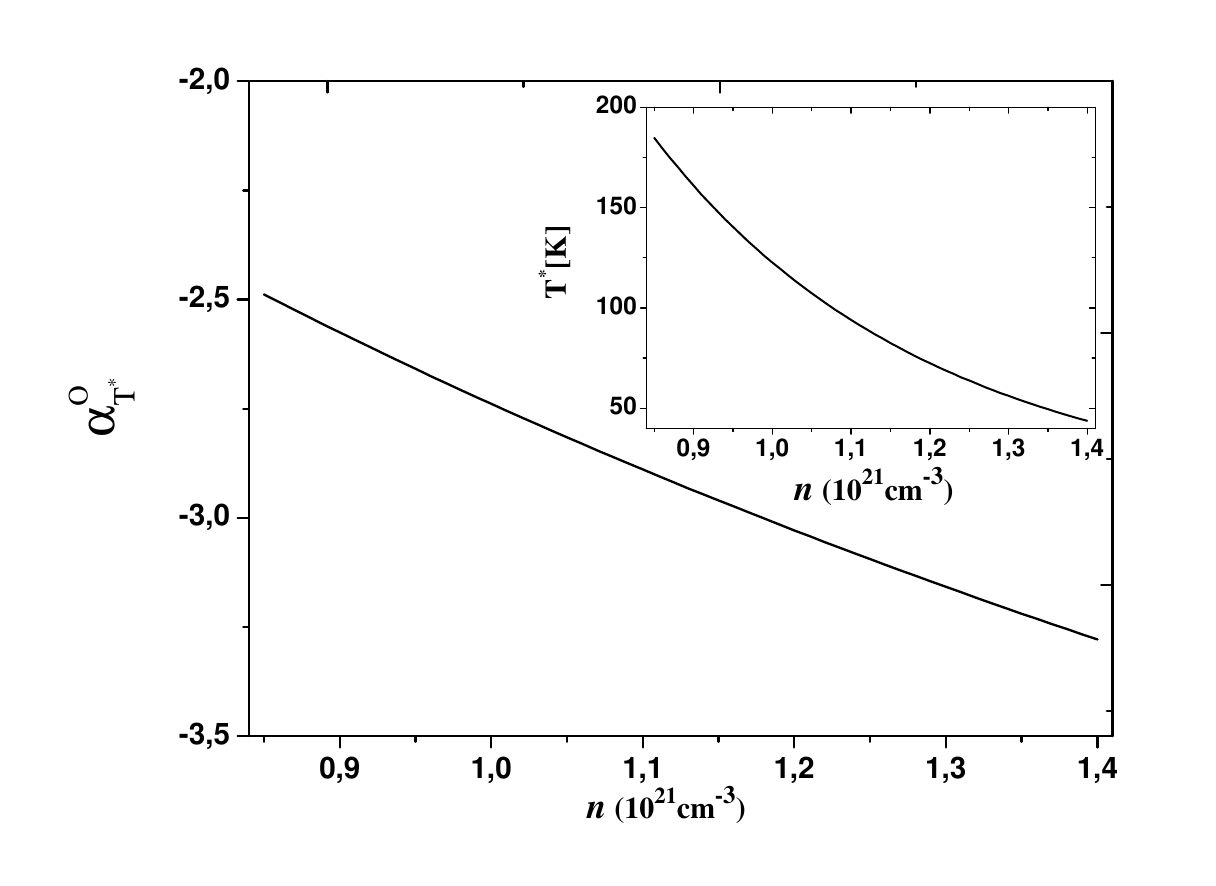}
\caption{\label{fig3:epsart}\footnotesize The doping dependence of
$\alpha^{O}_{T^{\ast}}$ (main figure) and $T^{\ast}$ (inset) for
$\lambda_{ph}=0.975$, $\lambda_{c}=0.820$ and
$\tilde{\varepsilon}=5.088$.}
\end{figure}
\begin{figure}
\includegraphics[width=0.51\textwidth]{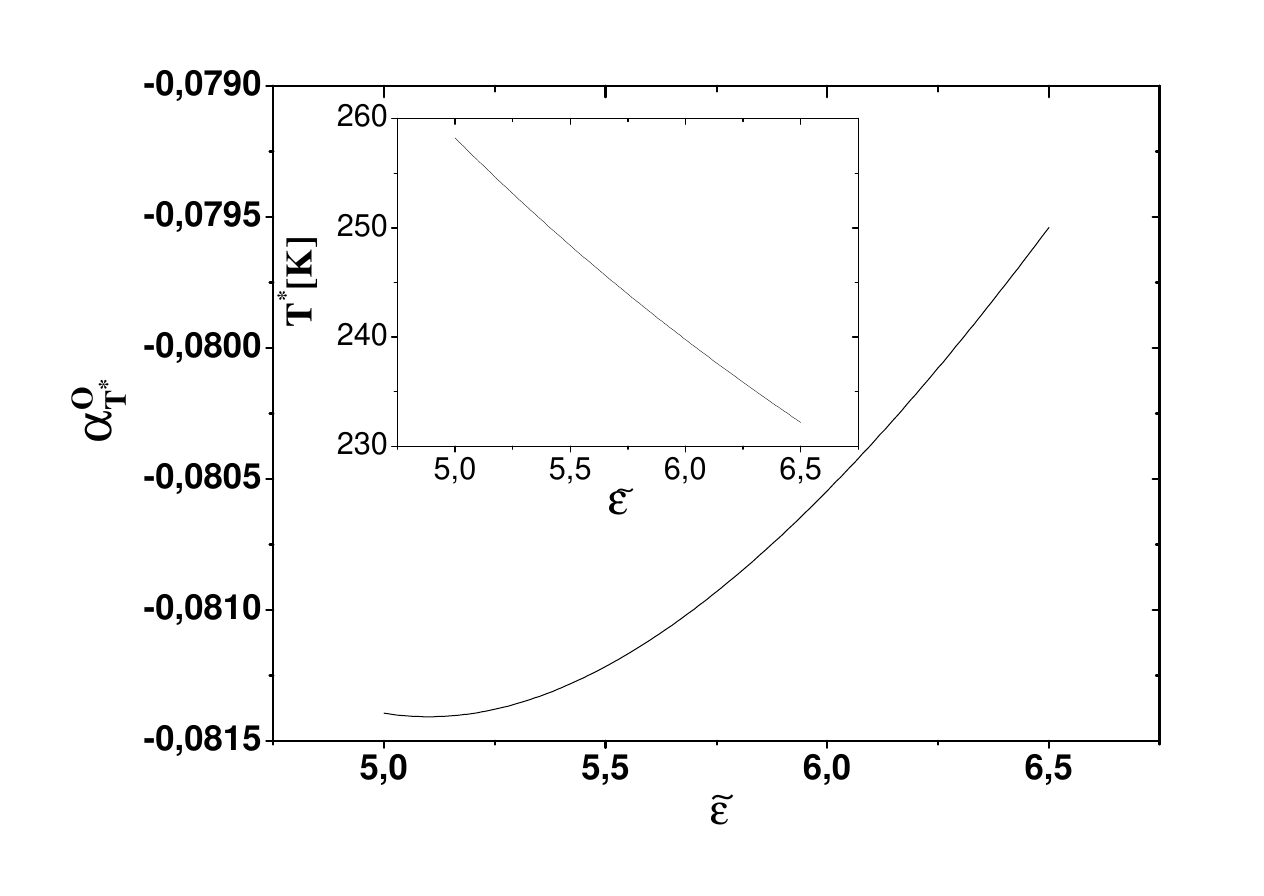}
\caption{\label{fig4:epsart}\footnotesize Variation of
$\alpha^{O}_{T^{\ast}}$ (main figure) and $T^{\ast}$ (inset) as a
function of $\tilde{\varepsilon}$ for $\lambda_{ph}-\lambda_{c}=0.3$
and $n=0.6\cdot10^{21} cm^{-3}$.}
\end{figure}
\par Our results provide a consistent picture of the existence of
crossover temperature $T^{\ast}$ above $T_{c}$ and peculiar isotope
effects on $T^{\ast}$ in cuprates. They explain why the small
positive (see Fig.1) and very large negative (see Figs.2 and 3)
oxygen isotope effects and the large negative and near-absent copper
isotope effects on $T^{\ast}$ are observed in various experiments.
The obtained $T^{\ast}$ is plotted in the insets of Figs. 1 - 3 as a
function of $\tilde{\varepsilon}$ and $n$ for different values of
$\lambda_{ph}$ and $\lambda_{c}$. The values of $\lambda^{\ast}$
varies from 0.3 to 0.5 and $T^{\ast}$ increases with decreasing $n$.
We have verified that $T^{\ast}$ decreases with increasing
$\tilde{\varepsilon}$ for $\lambda_{ph}<0.35$ and $\lambda_{c}<0.1$
(Fig.1). In contrast, $T^{\ast}$ increase with increasing
$\tilde{\varepsilon}$ for $\lambda_{ph}>0.4$ and $\lambda_{c}>0.25$
(Fig.2). The existing experimental data on $T^{\ast}$ and
$\alpha_{T^{\ast}}$ could be fitted with an excellent agreement
using Eqs. (\ref{Eq.5}) and (\ref{Eq.6}), and adjusting the
parameters $\hbar\omega_{LO}$, $\tilde{\varepsilon}$, $n$,
$\lambda_{ph}$ and $\lambda_{c}$ for each cuprate supercondutor. If
we choose $\hbar\omega_{LO}=0.05 eV$, $n=0.92\cdot10^{21} cm^{-3}$,
$\tilde{\varepsilon}=4.841 - 5.045$, $\lambda_{ph}\simeq0.297$ and
$\lambda_{c}\simeq0.077$, we see that $T^{\ast}=150 - 161 K$ and
$\alpha^{O}_{T^{\ast}}$ is very small (i.e.,
$\alpha^{O}_{T^{\ast}}=(0.0058 - 0.0098)<0.01$), which are
consistent with the experimental data of Ref. \cite{9}. Further,
using other sets of parameters $n=0.94\cdot10^{21} cm^{-3}$,
$\tilde{\varepsilon}=5.904 - 6.119$, $\lambda_{ph}=0.311 - 0.313$
and $\lambda_{c}= 0.067 - 0.070$, we obtain $T^{\ast}\approx150K$
and $\alpha^{O}_{T^{\ast}}\simeq0.0525 - 0.0623$, which are in good
agreement with the measured values: $T^{\ast}=150 K$ and
$\alpha^{O}_{T^{\ast}}=0.052 - 0.061$ for $\rm{YBa_2Cu_{4}O_{8}}$
(with $T_{c}=81 K$) \cite{10}. Figure 1 illustrates the predicted
behaviors of $\alpha^{O}_{T^{\ast}}$ and $T^{\ast}$ as a function of
$\tilde{\varepsilon}$ for $\lambda_{ph}<0.4$ and $\lambda_{c}<0.1$.
We see that $\alpha^{O}_{T^{\ast}}$ decreases slowly with decreasing
$\tilde{\varepsilon}$. Relatively strong electron-phonon and Coulomb
interactions (i.e., $\lambda_{ph}>0.5$ and $\lambda_{c}>0.5$) change
the picture significantly and cause $\alpha^{O}_{T^{\ast}}$ to
decrease rapidly with decreasing $\tilde{\varepsilon}$. In this case
the value of $\alpha^{O}_{T^{\ast}}$  is negative and becomes very
large negative with decreasing $\tilde{\varepsilon}$ or $T^{\ast}$.
The pictures shown in Figs.2 and 3 are likely realized in some
cuprates (which exhibit a large negative isotope exponent
$\alpha^{O}_{T^{\ast}}$) and explain another important puzzle of the
cuprates \cite{11}: the huge oxygen-isotope effect on $T^{\ast}$
observed in $\rm{HoBa_{2}Cu_{4}O_{8}}$, whose characteristic PG
temperature $T^{\ast}$ increases significantly upon replacing
$\rm{^{16}O}$ by $\rm{^{18}O}$. Indeed, with fitting parameters,
$n=0.88\cdot10^{21}cm^{-3}$, $\tilde{\varepsilon}=5.088$,
$\lambda_{ph}=0.975$ and $\lambda_{c}=0.82$, one can explain the
observed experimental data of Ref. \cite{11}. In this case, we
obtain $T^{\ast}(^{16}O)\simeq170.2K$,
$T^{\ast}(^{18}O)\simeq220.2K$, $\Delta
T_{O}^{\ast}=T^{\ast}(^{18}O)-T^{\ast}(^{16}O)\simeq50K$ and
$\alpha^{O}_{T^{\ast}}\simeq-2.54$, which are in remarkably good
agreement with the experimental data $T^{\ast}(^{16}O)\simeq170K$,
$T^{\ast}(^{18}O)\simeq220K$, $\Delta T_{O}^{\ast}\simeq50K$ and
$\alpha^{O}_{T^{\ast}}\simeq-2.2\pm0.6$ \cite{11}. The above
predicted behaviors of $T^{\ast}$ and $\alpha^{O}_{T^{\ast}}$ could
be checked experimentally in other slightly underdoped and optimally
doped cuprates. We have also performed similar calculations for the
copper isotope effect on $T^{\ast}$ in slightly underdoped
$\rm{HoBa_{2}Cu_{4}O_{8}}$ (where the electron-phonon and Coulomb
interactions seem to be much stronger than in
$\rm{YBa_{2}Cu_{4}O_{8}}$), and for the oxygen and copper isotope
effects on $T^{\ast}$ in optimally doped
$\rm{La_{1.81}Ho_{0.04}Sr_{0.15}CuO_{4}}$. In order to obtain the
values of $T^{\ast}\simeq160$ and $\simeq185 K$ observed accordingly
in copper unsubstitituted and substituted
$\rm{HoBa_{2}Cu_{4}O_{8}}$, we took $\hbar\omega_{LO}=0.05 eV$,
$n=0.9\cdot10^{21}cm^{-3}$, $\tilde{\varepsilon}=4.937$,
$\lambda_{ph}=2.196$, $\lambda_{c}=2.064$. Then we found
$T^{\ast}(^{63}Cu)\simeq160 K$, $T^{\ast}(^{65}Cu)\simeq184.7 K$,
$\Delta T_{Cu}^{\ast}=T^{\ast}(^{65}Cu)-T^{\ast}(^{63}Cu)\approx25K$
and $\alpha_{T^{\ast}}^{Cu}\simeq-4.9$ in accordance with
experimental findings $T^{\ast}(^{63}Cu)\approx160 K$,
$T^{\ast}(^{65}Cu)\approx185 K$ and
$\alpha_{T^{\ast}}^{Cu}\simeq-4.9$ \cite{13}. In the orthothombic
$\rm{La_{2-x}Sr_{x}CuO_{4}}$ the optimally doped level
($x\simeq0.15$) corresponds to the value $n=0.8\cdot10^{21}cm^{-3}$.
By taking $\hbar\omega_{LO}=0.045 eV$, $\tilde{\varepsilon}=5.681$,
$\lambda_{ph}=0.578$ and $\lambda_{c}=0.463$ for
$\rm{La_{1.81}Ho_{0.04}Sr_{0.15}CuO_{4}}$, we obtained
$T^{\ast}(^{16}O)=T^{\ast}(^{63}Cu)\simeq60 K$,
$T^{\ast}(^{18}O)\simeq69.3K$, $\Delta T^{\ast}_{O}\simeq9.3 K$,
$\alpha_{T^{\ast}}^{O}\simeq-1.3$, $T^{\ast}(^{65}{Cu})\simeq60.7
K$; $\Delta T_{Cu}^{\ast}\simeq0.7 K$, which agree with the
experimental data of Ref. \cite{14}.
\par The situation is different for the deeply underdoped cuprate with a
small Fermi surface. In this case, the polaronic effect is strong
enough and the condition $\varepsilon_{F}<E_{p}$ is satisfied.
Therefore, $E_{p}+\hbar\omega_{LO}$ and $\lambda^{\ast}$ in
Eq.(\ref{Eq.1}) would be replaced by $\varepsilon_{F}$ and
$\lambda^{\ast}=(V_{ph}-V_{c})N_{p}(\varepsilon_{F})$, respectively.
It is easy to verify numerically that, for
$\varepsilon_{F}/k_{B}T^{\ast}\gtrsim3.7$ (or
$\lambda^{\ast}\lesssim0.7$), the PG temperature $T^{\ast}$ can be
determined from the BCS-like relation
\begin{eqnarray}\label{Eq.7}
k_{B}T^{\ast}\simeq1.14\varepsilon_{F}\exp{(-1/\lambda^{\ast}(\mu))},
\end{eqnarray}
where
$\lambda^{\ast}(\mu)=(\lambda_{ph}-\lambda_{c})(1+b\mu^{1/4})$.
\par Using Eqs. (\ref{Eq.4}) and (\ref{Eq.7}), we obtain
\begin{eqnarray}
\alpha_{T^{\ast}}=\frac{b\mu^{\frac{1}{4}}}{4(1+\frac{M}{M^{'}})(1+b\mu^{1/4})}\left[1-\frac{1}{\lambda^{\ast}(\mu)}\right].
\end{eqnarray}
The dependences $\alpha^{O}_{T^{\ast}}(\tilde{\varepsilon})$ and
$T^{\ast}(\tilde{\varepsilon})$ for $\hbar\omega_{LO}=0.05 eV$,
$\lambda_{ph}-\lambda_{c}=0.3$ and $n=0.6\cdot10^{21}cm^{-3}$ are
presented in Fig.4. As seen from Fig.4, the $\alpha^{O}_{T^{\ast}}$
is rather small and negative. The $\alpha^{Cu}_{T^{\ast}}$ is also
negative and nearly four times smaller than $\alpha^{O}_{T^{\ast}}$.
\par In conclusion, we have developed two new BCS-based approaches extended
to the intermediate coupling regime to describe the precursor
pairing of large polarons above $T_{c}$ and the peculiar isotope
effects on the PG temperature $T^{\ast}$ in high-$T_{c}$ cuprates
with small and large Fermi surfaces. Our results show that the
oxygen isotope effect on $T^{\ast}$ in optimally doped cuprates with
moderate electron-phonon coupling strength ($\lambda_{ph}<0.4$) and
weak Coulomb repulsion ($\lambda_{c}<0.1$) is small positive. At
$\lambda_{ph}>0.5$ and $\lambda_{c}<0.5$ the negative oxygen and
copper isotope effects on $T^{\ast}$ in the same compound (e.g.,
$\rm{La_{1.81}Ho_{0.04}Sr_{0.15}CuO_{4}}$) is large and negligible,
respectively. While the very large negative oxygen and copper
isotope effects on $T^{\ast}$ exist in some cuprates with large
Fermi surfaces at $\lambda_{ph}<1$, $\lambda_{c}\lesssim0.8$ and
$\lambda_{ph}>\lambda_{c}>2$, respectively. The existing
experimental data and discrepancies between experiments measuring
the isotope effects on $T^{\ast}$ in various cuprates are
consistently explained by the proposed general BCS-like pairing
model. We predict that the isotope effects on $T^{\ast}$ in the
deeply underdoped cuprates are sizable and negative.

We thank B.Ya. Yavidov, E.M. Ibragimova and M. Ermamatov for useful
discussions. This work was supported by the Foundation of the
Fundamental Research of Uzbek Academy of Sciences, Grant No
FA-F2-F070+F075.

\end{document}